\documentclass[pra, 12 pt]{revtex4}
\usepackage[english]{babel}
\usepackage{amsmath}
\usepackage{epsfig}

\begin{document}

\title{FREQUENCY DEPENDING PERMITTIVITY OF THE COULOMB SYSTEM WITH BOSE-EINSTEIN CONDENSATE}

\author{V.B.Bobrov, S.A.Trigger}

\address{Joint\, Institute\, for\, High\, Temperatures, Russian\, Academy\,
of\, Sciences, 13/19, Izhorskaia Str., Moscow\, 125412, Russia;\\
emails:\, vic5907@mail.ru,\;satron@mail.ru}

\begin{abstract}

 The second-order singularity is found in the low-frequency region of the permittivity of a homogeneous and isotropic system of charged particles consisting of electrons and boson nuclei. This singularity is caused by the existence of a Bose-Einstein condensate for nuclei. The result obtained leads to the existence of the "nuclei superconductivity", which can be experimentally verified in superfluid He II. The results of the proposed an experiment can be considered as a direct proof of the existence of a Bose-Einstein condensate in superfluid He II. \\

PACS number(s): 05.30.Jp, 52.25.Mq, 03.75.Kk, 03.75.Nt\\

\end{abstract}

\maketitle

\section{Introduction}

The experimental detection of a Bose-Einstein condensate (BEC) in rarefied gases of alkali elements [1-3] not only opened a new field of research in the ultralow temperature region [4,5], but also was an indirect confirmation of the possibility of using the BEC concept in the microscopic theory of superfluid He II [6,7].
In the interpretation of these phenomena, it is a priori considered that the corresponding fluids consist of electrically neutral "atoms" as initial microparticles. In particular, BEC appearance in rarefied gases is associated with confinement of alkali element atoms in magnetic traps due to diamagnetism of these atoms. Magnetic traps are filled with rarefied gas by cooling it by a laser with a frequency lower than the absorption frequency of corresponding atoms (see [8] for more details). In this case, the interaction of laser radiation with atoms is described in terms of atomic polarizability within the dipole approximation (see [9] for more details).

In turn, superfluid He II atoms have zero dipole moment. Therefore, the experimental results [10-12] showing electrical activity of superfluid He II appeared rather unexpected. First of all, the case in point is observation of an electromagnetic field arising during the propagation of second-sound waves or forced oscillations of the normal component velocity. To date, a number of theoretical models are proposed to describe this effect (see [13-17] and references therein). No less interesting is superradiant Rayleigh scattering  from a BEC, detected in [18-20], which differs from Rayleigh scattering in fluids (see [21] and references therein).

We would like to pay attention that the consistent theoretical study of electromagnetic phenomena in medium implies the consideration of the atom as a composite particle consisting of  nucleus and electron [22]. This necessitates the study of neutral fluids as a Coulomb system (CS) which is a non-relativistic system of charged electrons and nuclei interacting with each other according to the Coulomb law in the presence of BEC for nuclei [23]. The concept of the neutral fluid as a CS took place in [14,15] when considering the electrical activity in He II.
Therefore, the difference between the CS and plasma concepts should be emphasized. Plasma is a specific CS in which the average density of the number of electrons in delocalized states (scattering states) is comparable to the average density of all electrons in a material under consideration [24]. In turn, the neutral fluid is a CS in which the average density of delocalized electrons is extremely low; hence, such a system in many cases can be considered in the beginning as a system of  atoms (see [25,26] and references therein). Within the adiabatic approximation for a subsystem of nuclei, the initial atom can be considered as a nucleus with electronic states localized near it. In this case, the size of the initial atom, defined by the distribution of the inhomogeneous electron density near the nucleus of the corresponding atom should be much smaller than the average distance between initial atoms [26]. However, the use of the adiabatic approximation leads to ambiguity when determining the pair interaction potential between initial atoms. The matter is that such an approach, in addition to the pair interaction between atoms, is associated with the three-particle interaction and the interaction of large number of atoms (see, e.g., [27]) whose effect cannot be reliably estimated. At the same time, as shown in [28], the form of the pair interaction potential of atoms is of fundamental importance for describing the quantum neutral fluid with a BEC. This leads to the necessity to develop of a corresponding theory based on the Coulomb model of matter to define the concept of the effective interaction potential of atoms in the fluid (see, e.g., [29]).

An alternative version widely used in the plasma theory is based on the equivalent consideration of electrons and nuclei in the CS without the adiabatic approximation for the subsystem of nuclei. To describe the CS with a BEC, this approach is in the early development stage [23], although the study of the so-called charged Bose gas which is a model single-component system of charged bosons in a compensating background has a long history [30]. However, to describe the neutral fluids in which nuclei can be considered as charged bosons, the charged Bose gas model cannot be used due to the necessity of considering the strong interaction of electrons and nuclei, which results in the formation of initial atoms.
As a result, we come to the conclusion that the Coulomb model of matter should be applied to the consistent consideration of a BEC. In this regard, we proceed from the fact that electrons and nuclei in the CS are elementary particles within the used non-relativistic approximation. In this case, it is clear that the spin-independent interaction between charged particles cannot change the statistics of these particles. From this point of view, electrons as fermions are not involved in the BEC formation. The problem of a BEC for so-called composite particles, in particular, initial atoms or Cooper pairs for fermions, is not discussed in the present paper. In part II, we consider the definition of BEC for nuclei in the CS with boson nuclei. On this basis, in part III, we study the features of the frequency-dependent permittivity of the CS in the presence of a BEC for nuclei.

\section{Bose-Einstein condensate for nuclei}

When considering the non-relativistic CS, we suppose that nuclei (subscript $c$) are bosons, which results in the formation of a BEC for nuclei at low temperatures.

According to the general definition proposed by Penrose and Onsager [31], the existence of a BEC is associated with the anomalous spatial behavior of the equilibrium one-particle density matrix, which was called the off-diagonal long-range order (ODLRO) [32]. This statement for the homogeneous and isotropic CS in which the one-particle density matrix of nuclei has the form
$\gamma_c(\textbf{r},\textbf{r}')=\gamma_c(\mid\textbf{r}-\textbf{r}'\mid)$ is written as
\begin{eqnarray}
\lim_{\mid\textbf{r}-\textbf{r}'\mid\rightarrow\infty}\gamma_c(\textbf{r},\textbf{r}')=n_c^{BEC}\neq 0, \qquad \gamma_c(\textbf{r},\textbf{r}')=\langle\hat \Psi_c^+(\textbf{r})\hat \Psi_c(\textbf{r}')\rangle\label{A1}
\end{eqnarray}
where $n_c^{BEC}$  is the density of the number of nuclei in a BEC,  $\hat \Psi_c^+(\textbf{r})$ and $\hat \Psi_c(\textbf{r})$  are the field creation and annihilation operators for nuclei, respectively, angle brackets mean averaging over the Gibbs distribution. In the normal CS, $n_c^{BEC}=0$, i.e., BEC is absent. It is clear that the existence of a BEC in the CS when using definition (1) can be caused only by boson nuclei. Note, that the CS in the superconducting state of electrons is characterized by the existence of ODLRO for the two-particle electron density matrix [32].
As it is well known,  thermodynamic equilibrium in the CS is provided by satisfying the quasi-neutrality condition (see [33] for more details) written as
\begin{eqnarray}
\sum_{a=e,c}z_a e n_a=0, \label{A2}
\end{eqnarray}
where $n_a=\langle\hat N_a\rangle/V$  is the average density for the number of particles of type a with charge $z_a e$   and mass $m_a$  in the volume $V$, $\hat N_a=\int d\textbf{r} \hat \Psi_a^+(\textbf{r})\hat \Psi_a(\textbf{r})$  is the operator of the total number of particles of type $a$; subscript $e$  corresponds to electrons.
It should be considered that averaging over the Gibbs distribution in the statistical theory corresponds to the thermodynamic equilibrium state only after transition to the thermodynamic limit $\langle\hat N_a\rangle\rightarrow\infty, V\rightarrow\infty, n_a=\langle\hat N_a\rangle/V=const$. This means that in calculating the average values, a system in a very large (macroscopic), but finite volume $V$ , should be initially considered, and then the transition to the thermodynamic limit should be performed  [34].
To pass to the thermodynamic limit in Eq. (1), the field operators $\hat \Psi_c^+(\textbf{r})$ and $\hat \Psi_c(\textbf{r})$  are written as
\begin{eqnarray}
\hat \Psi_c^+(\textbf{r})=\frac{1}{\sqrt V} \sum_s\sum_\textbf{p}\hat c^+_{\textbf{p},s}\exp(-i \textbf{p}\textbf{r}) , \qquad \hat \Psi_c(\textbf{r})=\frac{1}{\sqrt V} \sum_s\sum_\textbf{p}\hat c_{\textbf{p},s}\exp(i \textbf{p}\textbf{r})  \label{A3}
\end{eqnarray}
where $\hat c^+_{\textbf{p},s}$   and $\hat c_{\textbf{p},s}$  are the creation and annihilation operators, respectively, for nuclei with momentum $\textbf{p}$  and spin projection $s$. Without loss of generality, we further assume that nuclei have zero spin. Taking into account expansion (3), we can represent the one-particle density matrix for nuclei in the homogeneous and isotropic CS as the Fourier series
\begin{eqnarray}
\gamma_c(\mid \textbf{r}-\textbf{r}'\mid)=\frac{f_c^{(V)}(\textbf{p}=0)}{V}+\frac{1}{V}\sum_{ {\textbf{p} \neq 0}} f_c^{(V)}({\textbf{p}})\exp(i \textbf{p} (\textbf{r}-\textbf{r}')), \label{A4}
\end{eqnarray}
where $f_c^{(V)}(\textbf{p})$  is the average  occupation number of nuclei with momentum $\hbar\textbf{p}$  (or the single-particle distribution function over momenta), the subscript $(V)$  means that a given function corresponds to a system in a very large (macroscopic), but finite volume $V$. Taking into account (1) and (4), the average density of the number of nuclei in a BEC is written as
\begin{eqnarray}
n_c^{BEC}=\frac{f_c^{(V)}(\textbf{p}=0)}{V}. \label{A5}
\end{eqnarray}
Hence, the average occupation number of nuclei with zero momentum $\langle \hat N_0\rangle=f_c^{(V)}(\textbf{p}=0)=\langle \hat c^+_0 c_0\rangle=n_c$  is a macroscopic quantity which is exactly the definition of BEC.
As a result, after transition to the thermodynamic limit, the distribution function over momenta for nuclei in the presence of BEC has the form
\begin{eqnarray}
 f_c({\textbf{p}})=\langle \hat N_0\rangle \delta_{\textbf{p},0}+f^{(over)}_c(p)[1-\delta_{\textbf{p},0}]\label{A6}
\end{eqnarray}
where $f^{(over)}_c(p)$  is the single particle distribution function for nuclei in the "overcondensate"
state at $\textbf{p}\neq0$. In this case, the average density of the number of nuclei $n_c=\langle\hat N_c\rangle/V$  is given by
\begin{eqnarray}
n_c=\lim_{V\rightarrow\infty}\frac{1}{V}\sum_{\textbf{p}}f_c^{(V)}(\textbf{p})=n_c^{BEC}+n_c^{(over)}. \label{A7}
\end{eqnarray}
\begin{eqnarray}
n_c^{(over)}=\lim_{V\rightarrow\infty}\frac{1}{V}\sum_{\textbf{p}\neq 0}f_c^{(V)}(\textbf{p})=\int \frac{d^3 p}{(2\pi)^3}f^{(over)}_c(p) . \label{A8}
\end{eqnarray}
where $n_c^{(over)}$  is the density of the number of nuclei in overcondensate states.
Thus, to describe the equilibrium CS in the presence of BEC, it is first necessary to consider the initial system in a very large, but finite volume   and then, after separating singular terms corresponding to the macroscopic number of nuclei in a BEC, to pass to the thermodynamic limit. It is easy to verify that a similar statement takes place when considering the inhomogeneous system with a BEC [35].

\section{CS permittivity in the presence of BEC for nuclei}

As is known [36], electromagnetic properties of the homogeneous and isotropic CS under the influence of a weak electromagnetic field are completely defined by the permittivity tensor
\begin{eqnarray}
\varepsilon_{\alpha\beta}(q,\omega)=(\delta_{\alpha \beta}-\frac{q_\alpha q_\beta}{q^2})\varepsilon^{tr}(q,\omega)+ \frac{q_\alpha q_\beta}{q^2}\varepsilon^{l}(q,\omega), \label{A9}
\end{eqnarray}
where $\varepsilon^{tr}(q,\omega)$  and $\varepsilon^{l}(q,\omega)$  are, respectively, the transverse and longitudinal permittivities accounting for the spatial and frequency dispersion. In this case, in the long-wavelength limit ($q\rightarrow 0$)
\begin{eqnarray}
\lim_{q\rightarrow 0}\varepsilon^{tr}(q,\omega)= \lim_{q\rightarrow 0}\varepsilon^{l}(q,\omega) =\varepsilon(\omega), \qquad \lim_{q\rightarrow 0}\varepsilon_{\alpha\beta}(q,\omega)=\varepsilon(\omega)\delta_{\alpha \beta}. \label{A10}
\end{eqnarray}
Within the linear response theory, the function $\varepsilon(\omega)$  is defined by the relations (see [37] for more details)
\begin{eqnarray}
\varepsilon(\omega)=1-\frac{\omega_p^2}{\omega^2}-\frac{4\pi\varphi(\omega)}{\omega^2}, \qquad \varphi(\omega)=\int_0^\infty dt \exp(i\omega t) f_\varphi(t), \label{A11}
\end{eqnarray}
\begin{eqnarray}
f_\varphi(t)=-\frac{i}{3\hbar V}\langle[\hat I^\beta(t),\hat I^\beta(0)]\rangle, \qquad \lim_{t\rightarrow\infty}f_\varphi(t)=0 , \label{A12}
\end{eqnarray}
where $\omega_p$  is the plasma frequency defined by the total number of electrons and nuclei in the CS,
\begin{eqnarray}
\omega^2_p=\sum_a \omega^2_a, \qquad \omega^2_a=\frac{4\pi z_a^2 e^2 n_a}{m_a}, \label{A13}
\end{eqnarray}
$\omega_a$  is the plasma frequency for particles of type $a$, $\hat I^\beta=\sum_a z_a e \hat I_a^\beta$  is the operator of the total electric current, $\hat I_a^\beta$  is the operator of the total flux of the number of particles of type $a$,
\begin{eqnarray}
\hat I_a^\beta=-\frac{i\hbar}{2 m_a}\int d \textbf{r} \{\hat \Psi_a^+(\textbf{r})\nabla_{\textbf{r}\beta}\hat \Psi_a (\textbf{r})-\nabla_{\textbf{r}\beta}\hat \Psi_a^+(\textbf{r})\hat \Psi_a (\textbf{r})\}, \label{A14}
\end{eqnarray}
\begin{eqnarray}
\hat A(t)=\exp(i\hat H t)\hat A \exp(-i\hat H t), \label{A15}
\end{eqnarray}
and $\hat H$  is the exact Hamiltonian of the CS in the non-relativistic approximation. Hereafter, summation over repeated indices is meant. Relations (11) and (12) should be understood in the thermodynamic limit. In this case, the time correlation function $f_\varphi(t)$  is such that the time integral (12) defining $\varphi(\omega)$  at any real frequencies $\omega$  converges [38].
From (14) and (15), it immediately follows that
\begin{eqnarray}
\frac{d \hat I_a^\beta}{dt}=\frac{1}{m_a}\sum_{b\neq a} \int d \textbf{r}_1 d \textbf{r}_2 \nabla_{\textbf{r}_1 \beta} U_{a b}(\textbf{r}_1-\textbf{r}_2) \hat n_a (\textbf{r}_1) \hat n_b (\textbf{r}_2), \label{A16}
\end{eqnarray}
where  $U_{a b}(r)$ is the Coulomb interaction potential for charged particles of types  $a$ and $b$; $\hat n_a(\textbf{r}) =\hat\Psi_a^+(\textbf{r}) \hat \Psi_a(\textbf{r})$. Therefore, for the charged Bose gas which is a particular case of the one-component plasma (OCP) model, the permittivity $\varepsilon^{OCP}(\omega)$  , according to (11)-(16) is written as
\begin{eqnarray}
\varepsilon^{OCP}(\omega)=1-\frac{\omega_c^2}{\omega^2}, \label{A17}
\end{eqnarray}
Then we study the frequency dispersion of the permittivity $\varepsilon(\omega)$   in the static limit $\omega\rightarrow 0$  for the two-component CS which is an adequate model of the real matter consisting of electrons and nuclei of one chemical element.
For the normal CS, the function $\varepsilon(\omega)$  in the static limit $\omega\rightarrow 0$  has the singularity
\begin{eqnarray}
\varepsilon(\omega)\mid_{\omega\rightarrow 0}\rightarrow \frac{4\pi i \sigma_{st}}{\omega}, \label{A18}
\end{eqnarray}
where $\sigma_{st}=\lim_{\omega \rightarrow 0} \sigma(\omega)$  is the static conductivity which is nonzero at a nonzero temperature for all known materials. The classification of materials by the static conductivity (electrical conductivity) into "conductors" (high conductivity), "dielectrics" (low conductivity), and "semiconductors" (intermediate conductivity strongly depending on external conditions) is conditional. Therefore, the value of  $\sigma_{st}$  for all materials is finite, although it is lower for dielectrics than for conductors by many orders of magnitude.
Relation (18) is a consequence of the known general formula relating the dynamic conductivity  $ \sigma(\omega)$   to the permittivity  $\varepsilon (\omega)$   (see, e.g., [36])
\begin{eqnarray}
\varepsilon(\omega)=1+ \frac{4\pi i \sigma (\omega)}{\omega}. \label{A19}
\end{eqnarray}
	In this case, the static permittivity  $\varepsilon_{st}=\lim_{\omega \rightarrow 0} Re \varepsilon(\omega)$ , being a finite value, varies from large negative values for conductors to about unity for dielectrics.
To establish the correspondence between relation (11) based on the linear response theory and limit relation (18), we use the operator equation
\begin{eqnarray}
\frac{d \hat P_a^\beta}{dt}=\hat I_a^\beta, \qquad \hat P_a^\beta=\int d \textbf{r} r_\beta n_a(\textbf{r}), \label{A20}
\end{eqnarray}
Then, twice integrating in parts in definition (11) for function $\varphi(\omega)$  and taking into account  (20), we find for the permittivity $\varepsilon (\omega)$
\begin{eqnarray}
\varepsilon(\omega)=\varepsilon_p (\omega)+ \frac{4\pi i f_p(\infty)}{\omega}, \qquad \varepsilon_p (\omega)=1+4\pi\alpha(\omega)-\frac{1}{\omega^2}(\omega_p^2-\Omega_p^2). \label{A21}
\end{eqnarray}
\begin{eqnarray}
\alpha(\omega)=\int_0^\infty dt \exp(i\omega t) f_\alpha(t), \qquad f_\alpha(t)= f_p(t)-f_p(\infty),  \label{A22}
\end{eqnarray}
\begin{eqnarray}
f_p (t)=\frac{i}{3\hbar V}\langle[\hat P^\beta(t),\hat P^\beta(0)]\rangle, \qquad f_p(\infty)=\lim_{t\rightarrow\infty}f_p(t), \label{A23}
\end{eqnarray}
\begin{eqnarray}
\Omega_p^2=\frac{4 \pi i}{3\hbar V}\langle[\hat I^\beta(0),\hat P^\beta(0)]\rangle. \label{A24}
\end{eqnarray}
	Definitions (23), (24) should be understood in the thermodynamic limit. In this case, the transition to the limit $t\rightarrow\infty$  for the time correlation function $f_p(t)$  in calculating  $f_p(\infty)$ should be performed after passing to the thermodynamic limit (see [39] for more details). Let us pay attention that relations (21)-(24) are valid under the condition
\begin{eqnarray}
\lim_{t\rightarrow\infty}\frac{1}{ V}\langle[\hat I^\beta(t),\hat P^\beta(0)]\rangle=0, \qquad \lim_{t\rightarrow\infty}\frac{1}{ V}\langle[\hat P^\beta(t),\hat I^\beta(0)]\rangle=0, \label{A25}
\end{eqnarray}
which provide the fulfillment of the limit relation (18) [38].

According to the above consideration, to consider the possible BEC existence in the CS, the calculation of $\Omega_p^2$  (24) taking into account (3) requires the transition from the coordinate representation of the operators $\hat I_a^\beta$  and $\hat P_a^\beta$   in definitions (13) and (19) to the momentum representation (see, e.g., [40])
\begin{eqnarray}
\hat I_a^\beta=\sum_s\sum_{\textbf{p}\neq 0}\frac{\hbar p_\beta}{m_a}\hat a^+_{\textbf{p} s} \hat a_{\textbf{p}s},\qquad \hat P_a^\beta=i\sum_s\sum_{\textbf{p}}\hat a^+_{\textbf{p} s}\nabla_{p \beta} \hat a_{\textbf{p}s}  \label{A26}
\end{eqnarray}
	Using the commutation relations for the creation and annihilation operators, from (26)  we find
\begin{eqnarray}
\langle[\hat I_a^\beta(0),\hat P_a^\beta(0)]\rangle=-3i\hbar\sum_s\sum_{\textbf{p}\neq 0}f_a^{(V)}({\textbf{p},s}), \qquad f_a^{(V)}({\textbf{p},s})=\langle\hat a^+_{\textbf{p} s} \hat a_{\textbf{p}s} \rangle^{(V)} \label{A27}
\end{eqnarray}
	Substituting (27) into (24) and taking into account (7), (8), we find
\begin{eqnarray}
\Omega_p^2=\omega_p^2-\omega^2_{BEC}, \qquad \omega^2_{BEC}=\frac{4\pi z^2e^2 n^{BEC}}{m_c} \label{A28}
\end{eqnarray}
	Thus, according to (21)-(24),
\begin{eqnarray}
\varepsilon(\omega)=1+4\pi\alpha(\omega)-\frac{ \omega^2_{BEC}}{\omega^2}+\frac{4\pi i f_p(\infty)}{\omega}. \label{A29}
\end{eqnarray}
	For the normal CS, $\omega^2_{BEC}=0$ ; hence, a comparison of (18) and (29) shows that
\begin{eqnarray}
\sigma_{st}=f_p(\infty). \label{A30}
\end{eqnarray}
(see [38] for more details). In the specific limiting state of "true insulator" for which $\sigma_{st}=0$, the permittivity $\varepsilon^{TI}(\omega)$  takes the form [41,42]
\begin{eqnarray}
\varepsilon^{TI}(\omega)=1+4\pi\alpha^{TI}(\omega). \label{A31}
\end{eqnarray}
so that the quantity $\alpha (\omega)$  has the meaning of material polarizability and has no singularities in the static limit $\omega\rightarrow 0$  (see (22)).
	
Let us now consider the quantity   (29) in the presence of BEC for nuclei  $\omega^2_{BEC}\neq0$. In this case, a higher-order singularity appears in the permittivity $\varepsilon (\omega)$  at  $\omega\rightarrow 0$ in addition to singularity (18). In other words, according to (29), in the presence of BEC for nuclei, we have
\begin{eqnarray}
\varepsilon(\omega)\mid_{\omega\rightarrow 0}\rightarrow -\frac{\omega^2_{BEC}}{\omega^2}, \label{A32}
\end{eqnarray}
	It follows from relation (32), taking into account the Fresnel formulas (see, e.g., [43]), that an electromagnetic wave with an extremely low frequency  $\omega$, incident on the interface the CS with a BEC for nuclei, will be almost completely reflected from the surface.
In other words, a weak electromagnetic field will not penetrate the CS in the presence of BEC for nuclei.

Furthermore, if we proceed from the definition of the dynamic conductivity  $\sigma(\omega)$   as a proportionality factor between the electric current density $\textbf{J}(\textbf{k},\omega)$  and the electric field strength $\textbf{E}(\textbf{k},\omega)$  in the weak inhomogeneity limit  $q\rightarrow 0$ (see [36] for more details),
\begin{eqnarray}
\textbf{J}(\textbf{k},\omega)=\sigma(\omega)\textbf{E}(\textbf{k},\omega), \label{A33}
\end{eqnarray}
taking into account (19), (28), (32), we find in the limit  $\omega\rightarrow 0$ that
\begin{eqnarray}
\textbf{J}(\textbf{k},\omega)=i\frac{4\pi z^2e^2 n^{BEC}}{m_c\omega} \textbf{E}(\textbf{k},\omega), \label{A34}
\end{eqnarray}
If we now consider the relation between $\textbf{E}(\textbf{k},\omega)$  and the vector potential of the electromagnetic field $\textbf{A} (\textbf{k},\omega)$  for the limit $q\rightarrow 0$: $\textbf{E}(\textbf{k},\omega)=i\omega\textbf{A} (\textbf{k},\omega)/c$, where $c$ is the speed of light, we find
\begin{eqnarray}
\textbf{J}(\textbf{k},\omega)=-\frac{ z_c^2e^2 n^{BEC}}{m_c c} \textbf{A}(\textbf{k},\omega). \label{A35}
\end{eqnarray}
As it is easy to see, relation (35) is similar to the London's equation for the superconducting electron current [44].
According to (19), (29), (33)-(35), in the general case of arbitrary frequencies $\sigma(\omega)$, we can present the electric current density $\textbf{J}(\textbf{k},\omega)$  in a weak electromagnetic field defined by the vector potential  $\textbf{A}(\textbf{k},\omega)$ in the form
\begin{eqnarray}
\textbf{J}(\textbf{k},\omega)=\textbf{J}^{BEC}(\textbf{k},\omega)+\textbf{J}^{(over)}(\textbf{k},\omega),
 \label{A36}
\end{eqnarray}
\begin{eqnarray}
\textbf{J}^{BEC}(\textbf{k},\omega)=-\frac{ z_c^2e^2 n^{BEC}}{m_c c} \textbf{A}(\textbf{k},\omega),
 \label{A37}
\end{eqnarray}
\begin{eqnarray}
\textbf{J}^{(over)}(\textbf{k},\omega)=\frac{\{i\omega\sigma_{st}+\omega^2\alpha(\omega)\}}{c}\textbf{A}(\textbf{k},\omega),
 \label{A38}
\end{eqnarray}
Representation (36) is similar to the separation of the electric current density in the weakly inhomogeneous electromagnetic field into the superconducting and normal components in the "electronic superconductivity" theory (see, e.g., [45]). We note, that such classification is conditional, since the current density $\textbf{J}^{(over)}(\textbf{k},\omega)$ , as well as the quantities $\sigma_{st}$ and $\alpha(\omega)$, depend on the existence of BEC in the CS. It also follows from relations (36)-(38) that the superconductor behavior in a finite-frequency $\omega$  electromagnetic field will change significantly in comparison with the static case ($\omega\rightarrow 0$).

\section{Conclusion}

We come to the conclusion that the "superconductivity of nuclei" takes place in the homogeneous and isotropic CS in the presence of BEC for nuclei. As follows from the above consideration, the "superconductivity of nuclei" is not directly related to the static conductivity $\sigma_{st}$ (30). The last one is probably controlled by the interaction of electrons with nuclei in "overcondensate" states. Thus, the transition to the state of the "superconductivity of nuclei" is possible for conductors, as well as for semiconductors and dielectrics.
These statements are direct consequences of that the derivation of relations (32), (35) is in fact based only on the definition of the BEC concept via ODLRO for the one-particle density matrix without use of the perturbation theory on the interparticle interaction in the CS (in contrast to the electronic superconductivity theory). It is obvious that specific calculations of the values of $n^{BEC}$ , $\sigma_{st}$ , and $\alpha(\omega)$  requires the use of these or those approximate methods.

An appropriate object to experimentally confirm the above results is the superfluid He II. The BEC currently obtained in rarefied gases is strongly inhomogeneous and seems inappropriate for the "superconductivity of nuclei" straightforward observation.  It should also be emphasized that an experimental validation of the existence of the "superconductivity of nuclei" is of fundamental importance for solving the problem of the relation between the superfluidity phenomenon and the existence of BEC.

\section*{Acknowledgment}

This study was supported by Russian Science Foundation (project no. 14-19-01492).
The authors are thankful to G.J.F. van Heijst, A.M. Ignatov, A.G. Khrapak, A.A. Roukhadze, P.P.J.M. Schram and A.G. Zagorodny for the fruitful discussions.\\

\end{document}